# Monolithically integrated high-power narrow-bandwidth microdisk laser

JIANGLIN GUAN,[1,2] CHUNTAO LI,[1,2] RENGHONG GAO,[3,4] HAISU ZHANG,[1,2] JINTIAN LIN,[3,4,†] MINGHUI LI,[3,4] MIN WANG,[1,2] LINGLING QIAO,[3] LI DENG,[1,2] AND YA CHENG[1,2,3,5,6,7,*]

[1]*State Key Laboratory of Precision Spectroscopy, East China Normal University, Shanghai 200062, China*
[2]*The Extreme Optoelectromechanics Laboratory (XXL), School of Physics and Electronic Science, East China Normal University, Shanghai 200241, China*
[3]*State Key Laboratory of High Field Laser Physics and CAS Center for Excellence in Ultra-Intense Laser Science, Shanghai Institute of Optics and Fine Mechanics (SIOM), Chinese Academy of Sciences (CAS), Shanghai 201800, China*
[4]*Center of Materials Science and Optoelectronics Engineering, University of Chinese Academy of Sciences, Beijing 100049, China*
[5]*Collaborative Innovation Center of Extreme Optics, Shanxi University, Taiyuan 030006, China*
[6]*Collaborative Innovation Center of Light Manipulations and Applications, Shandong Normal University, Jinan 250358, China*
[7]*Shanghai Research Center for Quantum Sciences, Shanghai 201315, China*
[†]*E-mail: jintianlin@siom.ac.cn*
[*]*E-mail: ya.cheng@siom.ac.cn*

Date: Dec. 22, 2022

arXiv

**Integrated on-chip microdisk lasers have attracted great attention as a light source of compact size, low lasing threshold and narrow bandwidth. However, challenges remain unresolved in terms of single mode operation at high output power while maintaining the ultra-narrow bandwidth. In this work, we demonstrate monolithically integrated on-chip single-frequency microdisk lasers coupled with bus-waveguides fabricated by photolithography assisted chemo-mechanical etching. Owing to the high-Q factor of a polygon whispering gallery mode formed in the microdisk and long cavity lengths (e.g., 409 μm and 1 mm), a microdisk laser with a narrow linewidth of 0.11 MHz and a maximum output power of 62.1 μW has been achieved at room temperature.**

---

On-chip single-mode micro-lasers are one of the key elements of the next-generation optoelectronic devices for optical communication, detection, metrology, and computing [1-4]. Such lasers usually are built upon micro-resonators which can reinforce the stimulated emission in the gain mediums by efficient confinement of light fields. In general, resonant modes with high quality (Q) factors and low mode volumes are required to improve the laser quality aiming at low threshold of pump power and high coherence [1-3]. Whispering-gallery-mode (WGM) micro-resonators featuring ultra-high quality (Q) factors and small mode volumes are the promising platform for high-performance micro-lasers [2,3,5-10]. A large variety of geometrical configurations such as microrings, racetrack micro-resonators, microspheres, microtoroids, and microdisks, have been developed to achieve single-mode microlasers [2,3], among which on-chip microdisk lasers [6] have shown unique advantages of high Q factors, possibility of dense integration [10], and capability of mode regulation by tailoring wedge angles of side walls of the microdisks [11,12]. So far, single-mode microdisk lasers have been demonstrated in individual microdisk cavities mainly by reducing the cavity size or introducing nano-structures such as gratings, slots, or nano-antennas for side-mode suppression, giving rise to single-mode lasing at various laser wavelengths ranging from the ultraviolet to the infrared [5]. However, the strategy of reducing cavity sizes will inevitably induce large bend losses, and that of using nano-structures for modes selection suffers from strong scattering loss, both of which make it difficult to maintain low lasing threshold and narrow laser linewidth. In addition, fast wavelength tunability is frequently required together with the narrow-bandwidth laser operation which relies on the electro-optic (EO) coefficient of the gain material as EO tuning provides the highest tuning speed as compared with the thermo-optic and acousto-optic tuning.

Recently, thin-film lithium niobate on insulator (LNOI) has shown great potentials as one of the most important platforms for high-performance photonic integrated circuits due to its excellent optical properties in terms of broad transparency window, large second-order nonlinear optical coefficient/linear electro-optic coefficient, and high refractive index contrast [13-22]. Efficient nonlinear optical processes and high-speed modulation with unprecedented performances have been demonstrated on LNOI platform. Moreover, optical gain has been introduced into LNOI platforms by doping lithium niobate with rare-earth elements such as erbium and ytterbium for lasing and optical amplification [22-38]. Until now, the highest reported maximum output power, to the best of our knowledge, is realized in rare-earth doped micro-lasers which reached several μW [23-34]. Generation of high-power single-frequency micro-lasers cannot be achieved simply by increasing the mode volumes in large microdisk cavities as multi-mode lasers will naturally be generated due to the small free spectral range within the optical gain bandwidth. The solution to this challenge requires an elegant control on the modes formed within a large microdisk by breaking the centrosymmetry of the microdisk cavities.

In this article, we demonstrated a monolithically integrated high-power narrow-bandwidth single-frequency microdisk laser using large-size microdisks fabricated on erbium-doped LNOI platform. The microdisk is monolithically integrated with a ridge waveguide to enable side-coupling of the pump laser into the microdisk and the generated single-mode laser out of the microdisk. The ridge waveguide introduces a weak perturbation on the whispering gallery modes (WGMs), enabling the formation of polygon modes with large free spectral range (FSR) [31,39,40] to avoid multi-mode lasing in such large microdisks. We measured a narrow linewidth of 0.11 MHz and a high output power of 62.1 μW from the large microdisk with a cavity length of 1 mm.

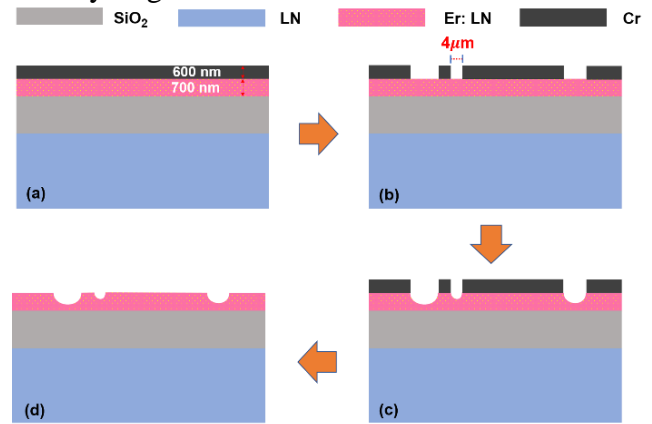

Figure 1. Illustration of the fabricated flow of the integrated microdisk laser. (a) Coating of a chromium (Cr) layer on top of the Er: LN. (b) Patterning the Cr film using space-selective femtosecond laser ablation. (d) Transferring the pattern of Cr film to Er: LN by chemo-mechanical etching. (d) Removing the residual Cr by wet chemical etching.

The erbium ion was doped in lithium niobate crystal with concentration of 0.1 mol.% during the crystal growth using Czochralski method. Then erbium doped lithium niobate crystal (Er: LN) was sliced into X-cut thin-film lithium niobate wafer with a thickness of 700 nm using smart-cut technique [41]. Here, the Er: LN thin film was supported by a silica layer with a thickness of 2 μm on a pure lithium niobate substrate, to form doped LNOI wafer. The un-suspended microdisks were fabricated on the LNOI wafer by photolithography assisted chemo-mechanical etching (PLACE) technique [42], which mainly consists of 4 steps, as illustrated in Fig. 1. First, a chromium (Cr) layer with a thickness of 600 nm was coated on the LNOI by magnetron sputtering. Second, the Cr layer was etched into patterns comprising microdisks side coupled with strips with gaps of 4 μm using space-selective femtosecond laser ablation. Third, the patterns were transferred into the thin-film lithium niobate by etching the exposed Er: LN thin film with a depth of 300 nm via chemo-mechanical etching. Finally, the residual Cr layer was removed in chemical etched solution.

The optical micrograph of fabricated device is shown in Fig. 2(a). The un-suspended microdisk possesses a diameter of 409 μm, which is side coupled with a ridge waveguide. Here, the top width of the ridge waveguide is 1.3 μm, and the coupling gap between the ridge waveguide and the microdisk is 4.0 μm, as shown in the partially enlarged drawing in Fig. 2(b). The largest etching depth is ~300 nm. The scanning electron microscope (SEM) image of the coupling region is also shown in Fig. 2(c), where the sidewall of the ridge waveguide is pretty smooth. While there are some defects along the periphery of the microdisk, which should be suppressed in the future via secondary chemo-mechanical polishing [43] and high temperate annealing [44].

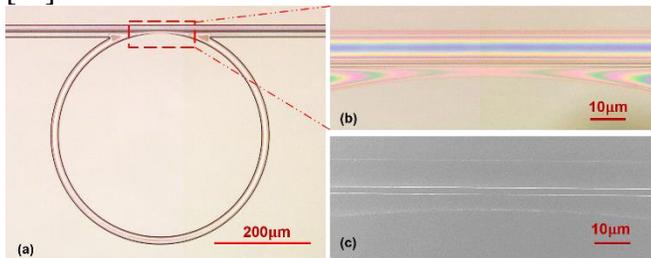

Figure 2. (a) Optical micrograph of the fabricated integrated microdisk laser. (b) Zoomed view of the coupling region of the integrated microdisk laser labeled in a box of (a). (c)Top-view scanning electron microscope (SEM) image of the coupling region of the integrated microdisk laser.

To achieve single-frequency lasing from polygon mode in the microdisk with a diameter of 409 μm, distributed feedback (DFB) laser diode with 980 nm wavelength was used as pump light. The linewidth of the pump light was about 0.5 nm. The pump light was coupled into the microdisk through the ridge waveguide using a lensed fiber. The polarization state of the pump light was adjusted by an inline polarization controller. The coupling efficiencies between the lensed fiber and the ridge waveguide was measured to be 10.8 % at 980 nm wavelength and 13.4 % at 1550 nm wavelength. The lasing signal was coupled out of the microdisk through the same ridge waveguide with another lensed fiber, and was sent into an optical spectrum analyzer (OSA) with a spectral resolution of 0.02 nm. Since the polygon pump mode and the polygon lasing mode possess almost the same spatial distribution which can be visualized by the up-conversion fluorescence of erbium ions [31,40], the intensity distribution of up-conversion fluorescence was captured with an optical microscope system mounted above the microdisk, consisting of an objective lens of numerical aperture of 0.42 and a charge coupled device (CCD).

When the pump power was increased to be higher than the threshold for lasing, single-frequency lasing signal was detected at 1552.82 nm wavelength, as shown in the spectrum in Fig. 3 (a). The captured up-conversion fluorescence emitted from the microdisk was shown in Fig. 3(b), exhibiting a hexagon pattern. Therefore, the side coupled ridge waveguide introduces weak perturbation into the circular microdisk to recombine WGMs to coherent polygon modes [31,39,40,45]. The output power as a function of the effective pump power is plotted in Fig. 3(c). The threshold was determined to be 825.39 μW from linear fitting. The total maximum output power of the microlaser was measured to be 39.7 μW when the pump power dropped in the microdisk was ~25.33 mW, which has been deduced considering the facet coupling losses. The conversion efficiency of the microdisk laser was $1.62 \times 10^{-3}$.

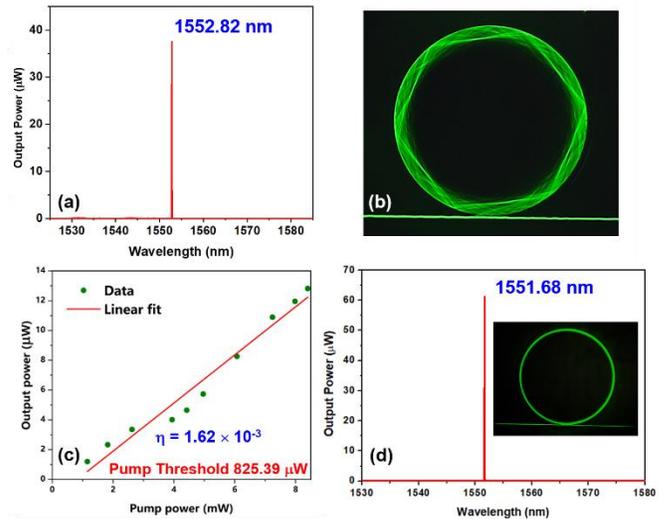

Figure 3. (a) Spectrum of the integrated microdisk laser with 409 μm diameter, exhibiting a single-frequency lasing at 1552.82 nm. (b) The green upconversion fluorescence of the integrated microdisk laser, showing a hexagon pattern when the 980 nm pump diode laser is injected. (c) Laser output power versus pump power dropped to the cavity showing a pump threshold of 298.6 μW. (d) Spectrum of the integrated microdisk laser with 1 mm diameter, confirming a single-frequency lasing at 1551.68 nm wavelength.

The larger microdisk with the diameter of 1 mm was also used to generate single-frequency microlaser,

which is exemplified by the spectrum in Fig. 3(d). Since the microdisk was very large, the captured polygon modes are consisted of high-order polygons locating near the periphery of the microdisk and resembling the conventional WGMs, as shown in the inset of Fig. 3(d). The lasing threshold was determined to 511.3 μW. The total maximum output power reached 62.1 μW when the pump power dropped in the microdisk was 10.93 mW, benefiting from the high Q factor as shown below. The conversion efficiency of the microdisk laser was $5.96\times10^{-3}$, which can be improved to a higher value by co-doping with erbium and ytterbium ions with higher absorption cross section [28,33,34].

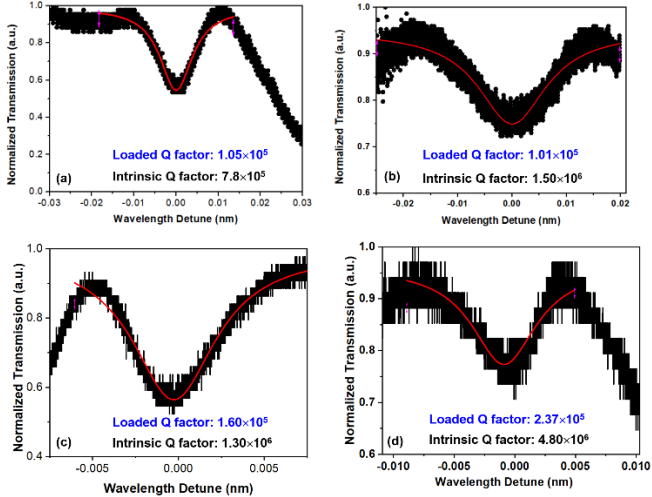

Fig.4. Transmission spectra of the integrated microdisks. (a) Lorentz fitting (red curve) reveals a loaded Q-factor of $1.05\times10^5$ at the wavelength of 980 nm. (b) Lorentz fitting (red curve) reveals a loaded Q-factor of $1.01\times10^5$ at the wavelength of 1552.82 nm. (c) Lorentz fitting (red curve) reveals a loaded Q-factor of $1.6\times10^5$ at the wavelength of 980 nm. (d) Lorentz fitting (red curve) reveals a loaded Q-factor of $2.37\times10^5$ at the wavelength of 1551.68 nm.

The Q factors of the pump mode and lasing mode of both microdisks were measured by using two wavelength tunable narrow-linewidth laser diodes (Model: TLB-6719 & 6728, New Focus Inc.) as pump lasers. The pump light was coupled into and out of the microdisk through the side-coupled ridge waveguide via the lensed fibers, and the transmission power was measured by a photo detector. The pump power dropped in the microdisk was set to a weak power of ~ 5 μW. Figures 4(a) and (b) show the transmission spectra of the waveguide coupled with the small disk around 980 nm and 1552.82 nm wavelength in the under-coupling state, respectively. The loaded Q factor of the pump mode at 980 nm was $1.05\times10^5$, from which the intrinsic Q factor was calculated to $7.8\times10^5$. The loaded Q factor of the lasing mode at 1552 nm was measured to be $1.01\times10^5$, from which the intrinsic Q factor was calculated to $1.5\times10^6$. The loaded Q factors of the pump mode and lasing mode in the large microdisk are shown in Figs. 4(c) and (d), respectively, where both the loaded Q factors are higher than $10^5$, and the intrinsic Q factors are higher than $10^6$. They are higher than their counterparts of the 409-um diameter microdisk because of the reduced bend loss.

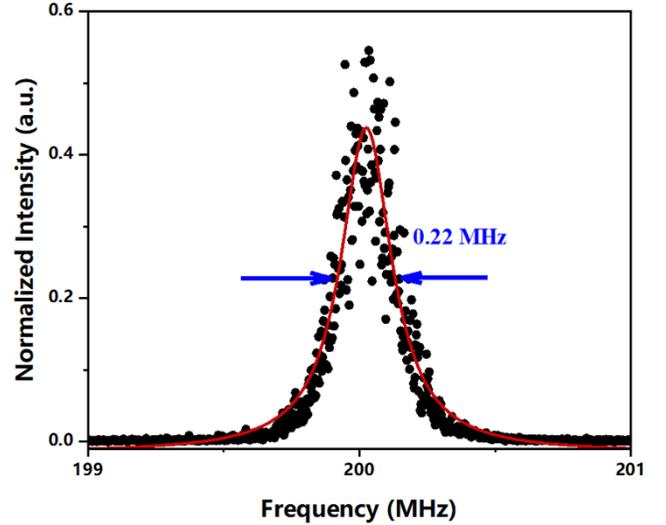

Fig.5. The Lorentz fitting (red curve) of the detected beating signal (black dots) featuring a laser linewidth of 0.11 MHz.

The linewidth of the microlaser in the larger microdisk was measured using an optical delayed self-heterodyne interferometer [1,29], which consists of two arms of an unbalanced interferometer. A single-mode fiber with a length of 5 km connected with an inline polarization controller was used as one arm, and the other arm was formed by an acoustic-optic modulator which shifted the light frequency by 200 MHz. The output power of the microlaser was first amplified to 1 mW by an erbium-doped-fiber amplifier. Then the signal was sent into a fast photo detector with a bandwidth of 1 GHz through the optical delayed self-heterodyne interferometer to generate a beating signal. A real-time electrical spectrum analyzer was used to analyze the beating signal. The detected signal was detected with a Lorentz shape with a central frequency of 200 MHz, as shown in Fig. 5. A Lorentz fitting curve is also plotted. The calculated linewidth is 0.22 MHz/2 = 0.11 MHz. To access a much narrower linewidth reported on suspended microdisk [31], the Q factors of the modes should be increased through optimizing the fabrication process of the integrated devices.

In conclusion, we have demonstrated a monolithically integrated microdisk laser operating at a single lasing frequency of narrow linewidth and tens of microwatts output power. The laser, once combined with efficient waveguide amplifier and EO tunability, can serve as miniaturized light sources for coherent optical communications, metrology, and lidar applications.